\documentclass[prb]{revtex4}% Physical Review B

\usepackage{graphicx}% Include figure files
\usepackage{dcolumn}% Align table columns on decimal point
\usepackage{bm}% bold math

\begin{document}

\preprint{APS/123-QED}

\title{Stochastic contribution to the growth factor in the ${\rm\Lambda}$CDM model}

\author{A.L.B. Ribeiro}
\affiliation{Center for Particle Astrophysics, FERMILAB, Batavia,IL 60510, USA \\ and
Departamento de Ci\^encias Exatas e
Tecnol\'ogicas, Universidade Estadual
de Santa Cruz, Ilh\'eus, 45650-000 BA, Brazil}
\email{albr@uesc.br}

\author{A.P.A. Andrade}
\affiliation{Departamento de Ci\^encias Exatas e
Tecnol\'ogicas, Universidade Estadual
de Santa Cruz, Ilh\'eus, 45650-000 BA, Brazil}
\email{apaula@uesc.br}

\author{P.S. Letelier}
\affiliation{ Departamento de Matem\'atica Aplicada-IMECC, Universidade Estadual de Campinas,
13081-970 Campinas, SP, Brazil}
  \email{letelier@ime.unicamp.br} 

\date{\today}% It is always \today, today,

\begin{abstract}
We study the
effect of noise on the evolution of the growth factor of density perturbations in the context of
the ${\rm\Lambda}$CDM  model. Stochasticity is introduced as a
Wiener process amplified by an intensity parameter $\alpha$.
By comparing the evolution of deterministic and stochastic cases for 
different values of $\alpha$ we estimate the intensity level
necessary to make noise relevant for cosmological tests
based on large-scale structure data. Our results indicate that the presence 
of random forces underlying the
fluid description can lead to significant deviations from the non-stochastic
solution at late times for $\alpha\ge 10^{-3}$.

\end{abstract}

%\keywords{cosmological parameters}%Use showkeys class option if keyword
\maketitle

\section{Introduction}

The problem of describing the growth of small density perturbations in the
universe consists of taking the differential equations of the fluid dynamics,
rewriting them as a unique wave equation for the density contrast, and finding
the solutions of this equation for different cosmologies. This is a classical 
problem of theoretical astrophysics, which began with the work of Jeans
 [1], its  general realtivistic generalization  was performed by Lifshitz [2].
 For a  modern presentation of this problem see for instance [3]. According to the traditional
approach, after the generation of the initial spectrum of density
perturbations in the early universe, the subsequent evolution of clustering is
deterministic, and does not admit a noise term in the dynamical equations.
However, stochasticity can be present due to processes hidden by
the  coarse-grained description of the fluid and whose typical time and
length scales are much shorter than those considered for large scale
structure formation (see e.g. [4]). Actually, noise
can be relevant in non-exceptional conditions, implying
that even if it is very weak in the beginning, its effects can be amplified
and have a non-negligible impact on the later
dynamical evolution of density perturbations [5,6].
At the same time, a very basic aspect of stochastic phenomena is related to
the differential law itself. It is well known that in a wide class of real 
dynamical situations, the relationship between random excitations and the response of a 
physical system is governed by differential equations which cannot be strictly deterministic,
but are stochastic (at a certain level) in nature (see e.g. [7]).
This suggests that exploring stochastic effects on the 
evolution of density perturbations can be very
important to modern cosmology, since the growth of large-scale structures is supposed to be used
to constrain the dark energy equation of state, see e.g. [8,9]. Therefore 
it seems important  to
understand if the presence of a stochastic term in the differential equation
still allows us to contrain cosmological models via 
large-scale structure data. In this work, we study the
effect of noise on the evolution of the growth factor in the context of
the ${\rm\Lambda}$CDM  model. Our aim is just to compare the evolution of
deterministic and stochastic cases and estimate if the typical differences
are similar or not to differences expected from assuming different
cosmological models.

\section{Gravitational Evolution of Stochastic Perturbations}

We consider a generic instability scenario in which perturbations are generated
by some mechanism in the early stages of the universe evolution, and start to
grow under gravity when non-relativistic matter begins to dominate the density
of the universe. The fluid is supposed to be pressureless and ideal, where
particles do not cross. The equations describing the motion of the fluid are
the continuity, Euler and Poisson equations. In Eulerian formulation they are
respectively
\begin{eqnarray}
{\partial\rho\over \partial t}+ 3{\dot{a}\over a}\rho + {1\over
  a}\nabla\cdot(\rho{\bf v}) &=& 0, \\
{\partial{\bf v}\over \partial t} + {1\over a}({\bf v} \cdot \nabla){\bf v} +
  {\dot{a}\over a}{\bf v} &=& -{1\over a}\nabla\phi, \\
\nabla^2\phi &=& 4\pi Ga^2 (\rho - \rho_b),
\end{eqnarray}
\noindent where $\rho_b$ is the homogeneous (\emph{background}) density of the fluid
and all the quantities are functions of the comoving coordinates ${\bf x} =
{\bf r} / a(t)$, with $a(t)$ given by the solution of the Friedmann
equation, see for instance [3]. If we define the density field as
\begin{equation}
  \rho({\bf x}, t)\equiv \rho_b(t)[1+\delta({\bf x},t)],
\end{equation}
\noindent where  $\delta({\bf x},t)$ is the density constrast,
 the fluid equations become
\begin{eqnarray}
{\partial\delta\over \partial t}+  {1\over
  a}\nabla\cdot(1+\delta){\bf v} &=& 0,\\
{\partial{\bf v}\over \partial t} + {1\over a}({\bf v} \cdot \nabla){\bf v} +
  {\dot{a}\over a}{\bf v} &=& -{1\over a}\nabla\phi, \\
\nabla^2\phi &=& 4\pi Ga^2 \delta.
\end{eqnarray}

\noindent Combining these equations, we find the second order differential equation for
$\delta$
\begin{equation}
{\partial^2\delta\over \partial t^2} + 
  2{\dot{a}\over a}{\partial\delta\over\partial t} - 4\pi G\rho_b\delta =  A,
\end{equation}
\noindent where
\begin{equation}
A= 4\pi G\rho_b\delta^2 + {1\over a^2}{\partial\delta\over \partial x^\alpha}
  {\partial\phi\over\partial x^\alpha} + {1\over a^2}
  {\partial^2\over\partial x^\alpha \partial x^\beta}(1+\delta)v^\alpha
  v^\beta.
\end{equation}

\noindent In the linear approximation, the term $A$ is not important and we simply have
\begin{equation} \label{eq:linear}
{\partial^2\delta\over \partial t^2} + 
  2{\dot{a}\over a}{\partial\delta\over\partial t} -
  4\pi G\rho_b\delta = 0,
\end{equation}
(see for instance [10]). In order to solve this equation, the
following decomposition is useful:
\begin{equation} \label{eq:decomp}
  \delta({\bf x},t)=\delta_s({\bf x})D(t).
\end{equation}
Since (\ref{eq:linear}) does not depend on spatial derivatives, $\delta$
evolves only in amplitude, preserving its original shape in the linear regime.
At the same time, the function $D(t)$  satisfies the following equation
\begin{equation}\label{eq:tevol}
  \ddot{D} + 2{\dot{a}\over a}\dot{D} - 4\pi G\rho_b D = 0,
\end{equation}
whose solution depends on the specific cosmological model adopted. This is the
standard way to find how small perturbations grow in the post-recombination
expanding universe. The fluctuations
$\delta({\bf x},t)$ clearly build a random field in space, since at a specific
position ${\bf x}_i$ we cannot know the exact value of $\delta$. In most
models $\delta_s({\bf x})$ is supposed to have a Gaussian distribution, while
$D(t)$ is always a deterministic function evolving with time according
to~(\ref{eq:tevol}). Actually, nothing is asserted about $D(t)$, which means
that there is an implicit assumption in~(\ref{eq:decomp}) saying that $D(t)$
does not have a stochastic nature. In this work, we assume instead that
noise can be present in the fluid due to the graininess of the underlying
physical system of particles [11]. For a wider discussion on
the possible origins of random forces in the large scale strucuture formation
see e.g. [4,5,6]. A simple way to introduce a stochastic term in~(\ref{eq:tevol}) is to suppose
that the field is submitted to a zero-mean randomly fluctuating frequency
$\zeta(t)$, such that
\begin{equation}\label{eq:tevols}
  \ddot{D} + 2{\dot{a}\over a}\dot{D} - [4\pi G\rho_b + \zeta(t)] D = 0,
\end{equation}
where $\zeta (t)=\alpha W$ is the magnification (by a factor $\alpha$) of a
Wiener process ($W$), a quantity which is supposed here to embrace all the possible
sources of random effects acting on the dynamical equation~(\ref{eq:tevol}).
Because of the stochastic nature of~(\ref{eq:tevols}), we are interested in the
statistical distribution of its solutions. This can be done directly by solving
the equation numerically many times with the same initial condition and
averaging over the results. The formal development of the theory of stochastic
differential equations can be found in [12], while
applications to stochastic phenomena in astrophysics are discussed
in [13]. 

\section{Stochastic Calculus }

In this section, we describe the method used to solve numerically the
stochastic differential equation (SDE)~(\ref{eq:tevols}). In short, we
transform this equation into a set of first-order equations, take their
integrals and iterate them from $t_i$ to $t_i + h$, then from there to $t_i + 2
h$, etc. For sufficiently small values of $h$, this approaches the atual
integral. The difference here is that instead of having usual integrals, we
must deal with It\^o (or Stratonovich) integrals, as described
in [14,15]. Suppose that we have to solve the following SDE:
\begin{equation}
 d\delta (t) = m[t, \delta (t)] \, dt + \sigma[t, \delta (t)] \xi(t) \, dt 
\end{equation}
where $ \delta(t = t_i) = \delta_i$ and $\xi(t)$ is the stochastic term. Here we assume that $\xi$ is the derivative
of a Wiener process $W(t)$. This is a continuous-time random walk with random jumps at every point in time, i.e.,
a step by step process in which the succession of steps is random and 
that in every step (jump) the variable changes values in a stochastic (random) way.
The Wiener process $W(t)$ is characterized by three facts: (i) $W(0)=0$; (ii) $W(t)$ is almost surely continuous;
(iii) $W(t)$ has independent increments for $0 \leq s < t$, e.g. [14].
Now, Equation (14) can then be re-written as
\begin{equation}
 d\delta (t) = m[t, \delta (t)] \, dt + \sigma[t, \delta (t)] \, d[W(t)].
\end{equation}
(This is far from being a trivial change, as the Wiener process may not admit a
time-derivative, but we ignore it here and follow the
very basic approach in the numerical procedure). Integrating both sides of the equation, we
obtain an \emph{ It\^o integral},
\begin{equation}
\delta(t) = \delta_i + \int_{t_i}^t m(s,\delta (s)) \, ds + \int_{t_i}^t
            \sigma(s,\delta (s)) \, d[W(s)]
\end{equation}
whose iterative counterpart in the interval $[t_j,t_{j+1}]$ is
\begin{eqnarray}
\delta (t_{j+1})  = \delta_j &+& \int_{t_j}^{t_{j+1}} m(s, \delta (s)) \, ds
                     \nonumber \\
                 &+& \int_{t_j}^{t_{j+1}} \sigma(s, \delta (s)) \,  d[W(s)].
                  \label{eq:itoiter}
\end{eqnarray}

\noindent Once in the form of an It\^o (or rather, for technical reasons, a Stratonovich)
integral, the equation is expanded about $t_i$ and solved with a fixed number
of terms in the expansion (here, we included terms of up to three nested
integrals),  see chapter 5 of [14] for  details.

\section{Growth Rates and Cosmology}

Observational evidence suggests that the universe is currently
accelerating, which would imply the existence of a significant unknown (dark)
energy component pervading the whole universe as
a homogeneous fluid [16,17,18,19]. To the moment, the existence of a 
cosmological constant seems to be the
most economical explanation for the present data, although we have no
fundamental physics solution to the coincidence and fine tuning problems, e.g. [20].
Trying to understand what is really behind the universe
acceleration, a diverse set of projects designed to probe the nature of
dark energy are in progress now. In fact, the increasing quantity and quality 
of large-scale structure data will soon allow us to use structure formation to constrain
the dark energy equation of state, e.g. [21].
Dark energy affects the growing mode of density perturbations through the damping
term $H=\dot{a}/a$ in Eq. (13), making structure formation an
interesting cosmological discriminator. However, differences in linear
growth rates (in dark energy models)
compared to a constant ${\rm\Lambda}$ for $z\lesssim 3$ are
so small (of order $10^{-2}$) that it may be extremely difficulty 
to rule them out observationally [22]. In this work we study the effect of a hypothetical 
stochastic contribution to the linear growth factor. Assuming it
as a Wiener process, we would like to know the intensity level
of this noise term to produce significant differences between 
stochastic and non-stochastic ${\rm\Lambda}$CDM model.
Accordingly, we take just the case of a flat universe with 
cosmological constant, so that

\begin{equation}
H^2={8\pi G\over 3}\rho_m +{\Lambda\over 3}.
\end{equation}

\begin{table}
  \begin{tabular}{|c|c|c|c|}
    \hline
            $\alpha$ & $\sigma_{f\!,\rm{a}}$ & $\sigma_{f\!,\rm{ns}}$ &
                        $|\Delta_{\rm{a},\rm{ns}}|$ \\
    \hline
            0.0001 & $1.070 \times 10^{-6}$ & $1.777 \times 10^{-6}$ &
                      $2.347 \times 10^{-5}$ \\
            0.001  & $3.230 \times 10^{-3}$ & $3.287 \times 10^{-3}$ &
                      $5.830 \times 10^{-2}$ \\
            0.005  & $6.040 \times 10^{-2}$ & $6.611 \times 10^{-2}$ &
                      $3.584 \times 10^{-1}$ \\
            0.01   & $3.200 \times 10^{+2}$ & $3.233 \times 10^{+2}$ &
                      $0.668 \times 10^{+1}$ \\
    \hline
  \end{tabular}
  \caption{Statistics for 200 functions in the entire time range. \label{tab:100its}}
\end{table}

\noindent Using our simulation code, we performed a number of simulations for the
case $\Omega_m = 0.3$, $\Omega_\Lambda = 0.7$ and $h = 0.7$. We carry out
 200 simulations for each of four values of $\alpha$: 0.0001,
0.001, 0.005 and 0.01, within the redshift range $z=1000\rightarrow 0$ or
$t=0\rightarrow 860$ in our simulations units. The range of the magnification
factor $\alpha$ is arbitrary, but chosen to be large enough to probe the 
levels at which the noise term turns important to the growth rates.
Also, our choices of $\alpha$ respect the fact that we do not expect large
signatures of the stochastic term near the last scattering surface due
to the high entropy per baryon by this epoch [23]. The values of $\alpha$
studied here just produce late time (significant) effects.
The quantity that we are interested in this analysis is the fractional
error with respect to the growth rate of the ${\rm \Lambda}$CDM model

\begin{equation}
f\equiv{\Delta D\over D}= {D_{stoch} - D_{\Lambda CDM}\over D_{\Lambda CDM}}
\end{equation}

\noindent Table~\ref{tab:100its} shows the effect of increasing
$\alpha$. It presents three basic quantities: $\sigma_{f\!,\rm{a}}$ is the largest
standard deviation of all functions when compared with the
average function; it indicates how much the functions spread around
the average.  $\sigma_{f\!,\rm{ns}}$ is the same, but compared to
the non-stochastic solution rather than the average. 
Finally, $|\Delta_{\rm{a},\rm{ns}}|$ is the absolute value of
the largest difference (for all $t$) between the average of 200
simulations and the non-stochastic solution. As expected, the maxima happen
at $z=0$ for all cases. In Figure 1, we present the evolution
of $\Delta D/D$ for each value of $\alpha$. Note that in all cases the
presence of the stochastic term is not important at early times, but
it becomes increasingly relevant at late times. We also see that the
intensity level of the Wiener process can drive the growth rate contrast
into a very noisy regime, with amplitudes higher than $10^{-2}$ for
$\alpha\geq 10^{-3}$.

\begin{figure}
\includegraphics[width=6.0cm,height=6.0cm,angle=0]{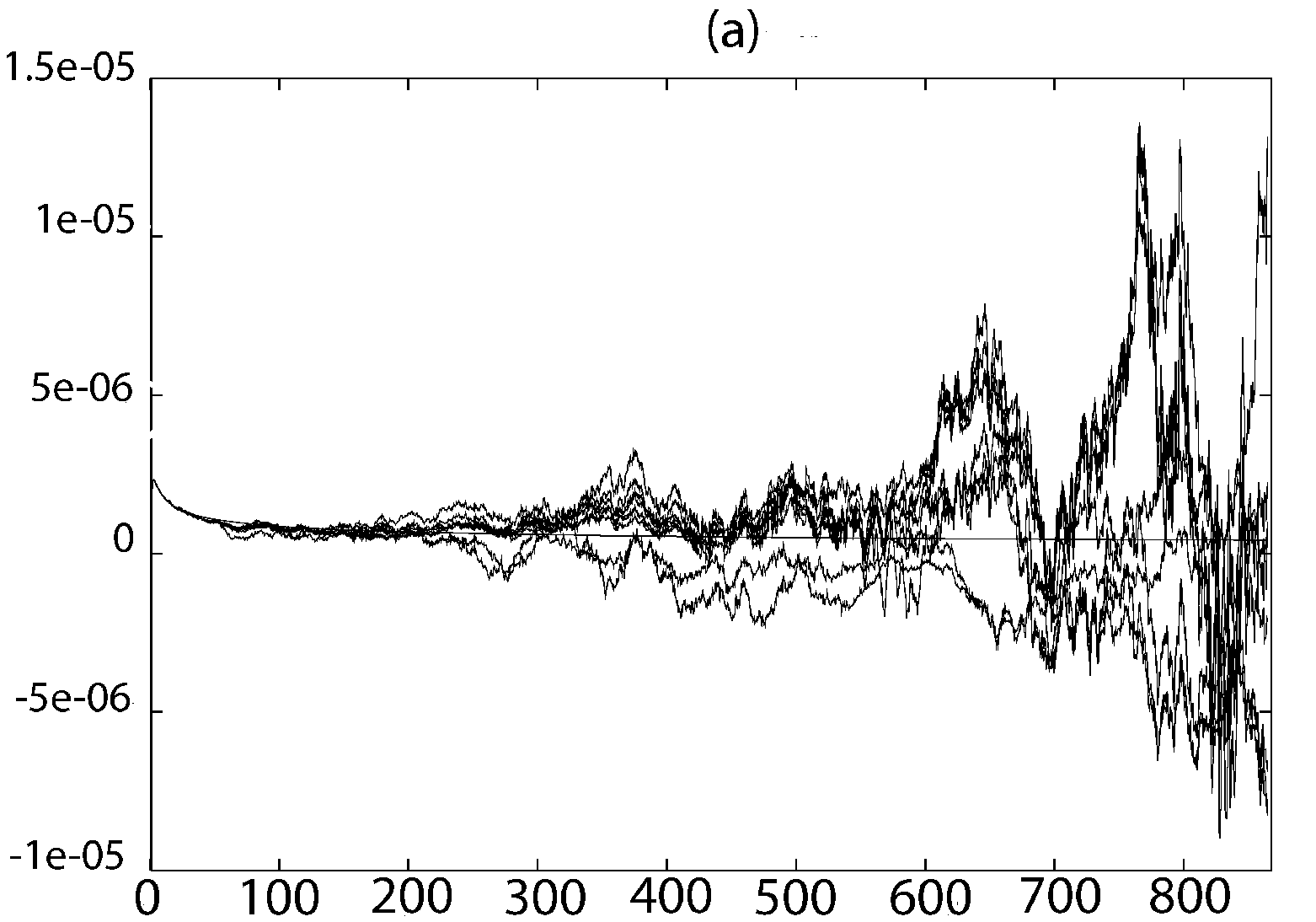}
\includegraphics[width=6.0cm,height=6.0cm,angle=0]{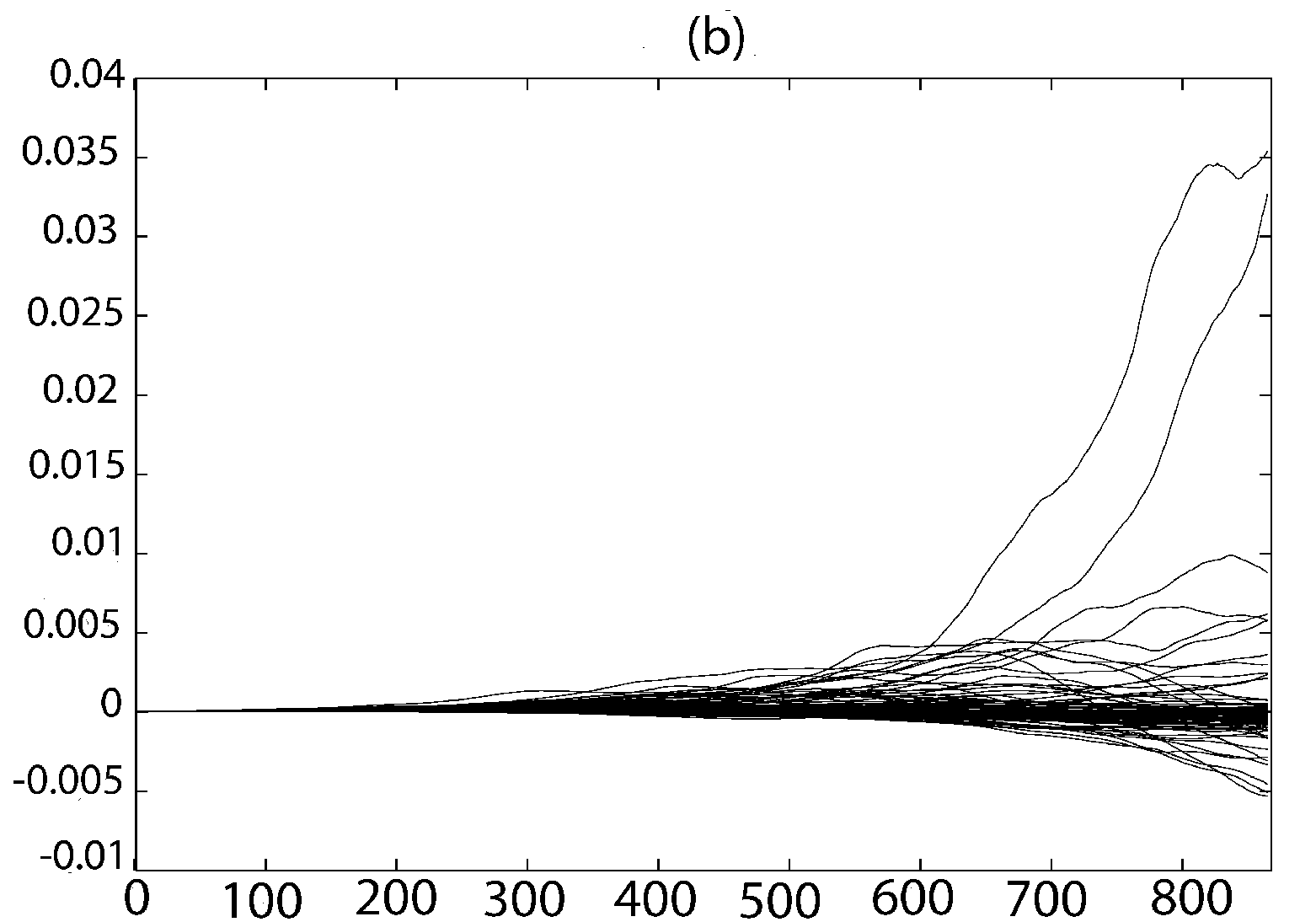}
\includegraphics[width=6.0cm,height=6.0cm,angle=0]{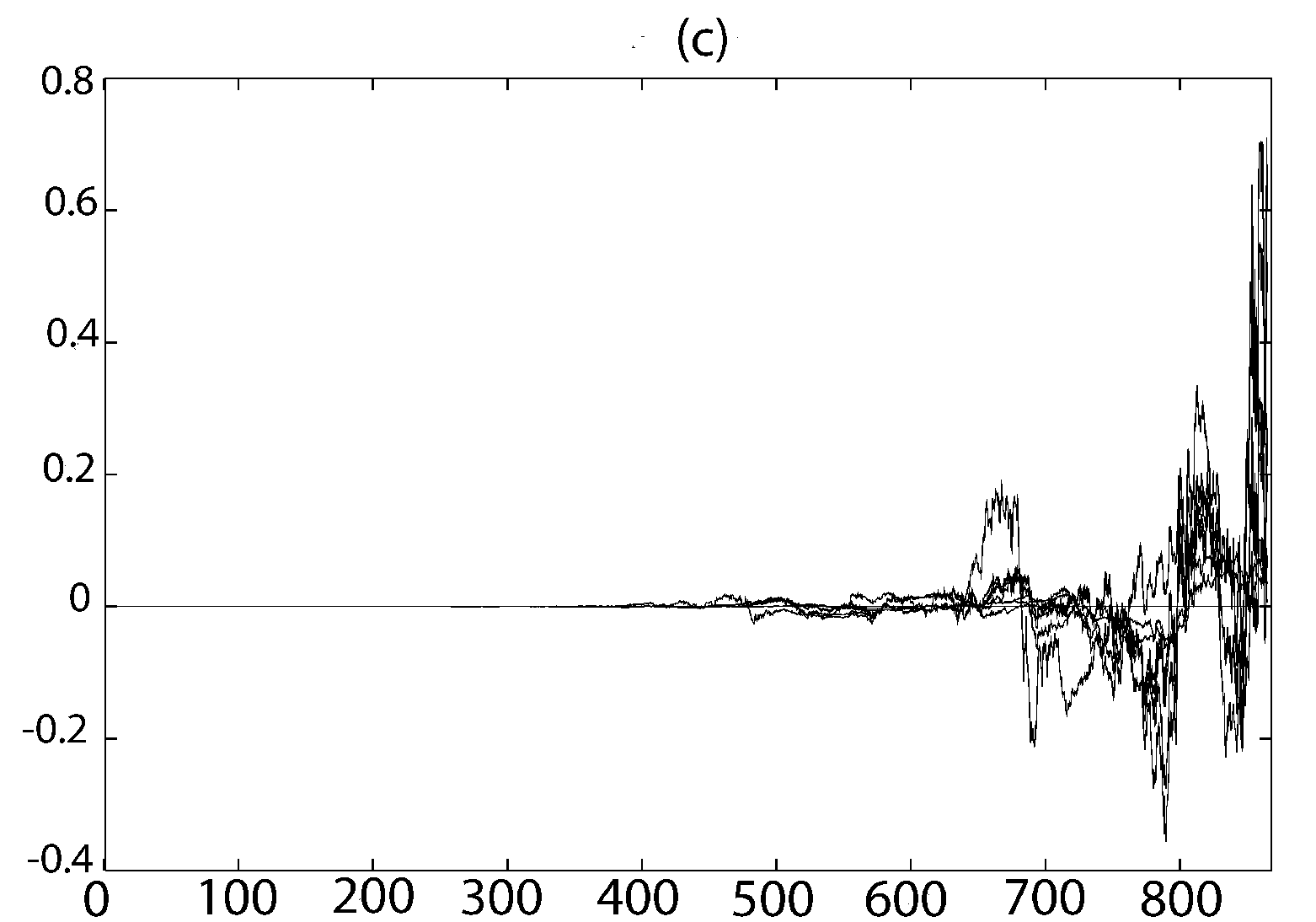}
\includegraphics[width=6.0cm,height=6.0cm,angle=0]{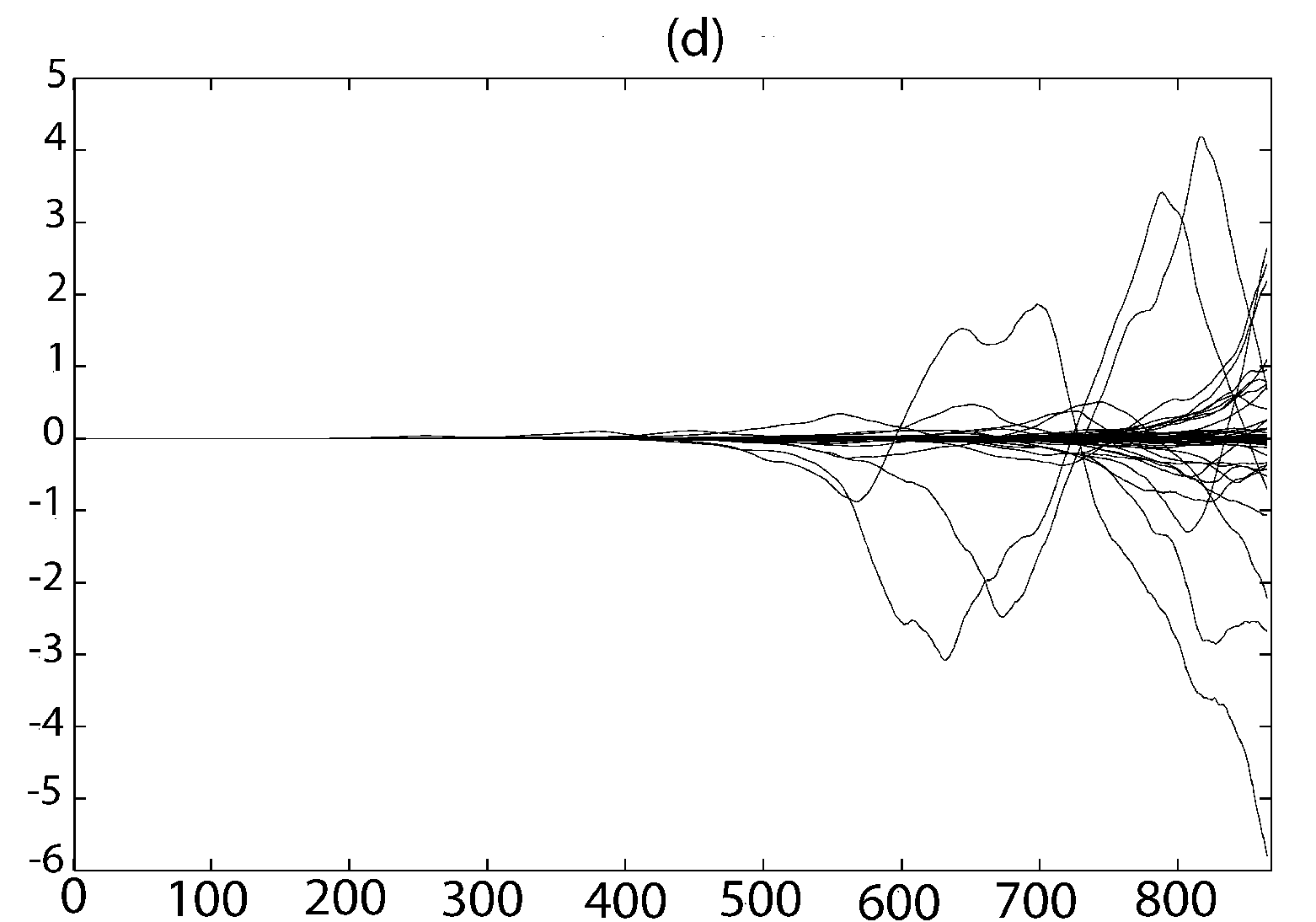}
%\subfigure{\epsfig{file=psDf1b.ps,angle=270,width=8cm}}\\ \subfigure{\epsfig{file=psDf1c.ps,angle=270,width=8cm}}
\caption{$\Delta D/D$ against t (simulation units) for $\alpha=0.0001,0.001,0.005,0.01$
(from top to bottom, following pannels (a) to (d)).}
\end{figure}

\section{Discussion} 

Our stochastic simulations suggest that the presence of random forces under the cosmic
fluid description can lead to significant deviations from the non-stochastic
solution for the growing mode at late times. In fact, 
for intensity levels of a
Wiener process higher than 
$10^{-3}$, the stochastic contribution to the growth factor would be
of the same order as the predictions of typical differences between ${\rm\Lambda}$CDM model
and dark energy models [22]. Hence, this noise contribution
would imply an additional difficulty to
rule out models using large-scale structure data.
In the present work, the introduction of stochasticity is completely {\it ad hoc}
and our toy model based on the amplification of a Wiener process
is quite arbitrary. However, our purpose is just to figure out if such
contribution could be relevant to cosmology even for a simple implementation
of the noise term. Actually, if noise should enter or not in the cosmic density
evolution description is still a matter of debate, see e.g. [4]. Our results
just raise the problem of distinguishing cosmological models using structure formation
in the context of a cosmic fluid with underlying random forces. Of course, to
pursue this effect more properly, theoretical efforts are necessary  in order to
understand the expected noise levels {\it ab initio} and the correct way to model and introduce
them in the fluid description. 

\begin{acknowledgments}
We thank the referee for useful suggestions.
A.L.B.R.  and P.S.L.  thank partial  support of CNPq. P.S.L. also thanks partial support of FAPESP. 
We thank Dr. F. Bonjour for his computational support.
\end{acknowledgments}

\thebibliography{1}
~~~~~~~~~~~~~~~~~~~~~
\eject
\noindent {\large{\bf References}}

~~~~~~~~~~~~~~~~~~~~

\bibitem{1} Jeans, J.H., Phil. Trans. Roy. Soc, 199, 1 (1902)
\bibitem{2} Lifshitz, E.M., J. Phys. Acad. Sci. (USSR), 10, 116 (1946)
\bibitem{3} Longair, M. in Galaxy Formation, Springer, Berlin (1998)
\bibitem{4} Buchert, T., Dominguez, A. and P\'erez-Mercader, J., Astronomy and Astrophysics, 349, 343 (1999)
\bibitem{5} Barbero, J.F., Dominguez, A., Goldman, T., and P\'erez-Mercader, J., Europhys. Lett, 38, 637 (1997)
\bibitem{6} Dominguez, A., Hochberg, D., Mart\'{\i}n-Garcia, J.M., 
P\'erez-Mercader, J. and Schulmann, L., Astronomy and Astrophysics, 344, 27 (1999)
\bibitem{7} Kloeden, E. and Pearson, R.A., J. Austr. Soc., B20, 1 (1977)
\bibitem{8} Linder, E.V. and Jenkins, A., MNRAS, 346, 573 (2003)
\bibitem{9} Benabed, K. and Bernardeau, F., Phys. Rev. D, 64, 083501 (2001)
\bibitem{10} Peebles, P.J.E., in Large Scale Structure of the Universe, Princeton University Press, 1980.
\bibitem{11} Lifshitz, E.M. and Pitaevskii, L.P., in Statistical Physics, Part 2
(Course of Theoretical Physics, Vol. 9) Pergamon Press, Oxford (1980)
\bibitem{12} Sobczyk, K., in Stochastic Differential Equations, Kluwer Academic Publishers, Dordrecht (1991)
\bibitem{13} Knobloch, E., Vistas in Astronomy, 24, 39 (1980)
\bibitem{14} Kloeden, E. and Platen, E., in Numerical Solution of Stochastic Differential Equations,
Springer, Berlin (1992)
\bibitem{15} Kloeden, E., Platen, E. and Schurz, H., in The Numerical Solution of Stochastic Differential Equations Through Computer Experiments, Springer, Berlin (1993)
\bibitem{16} Riess, A. et al., AJ, 116, 1009 (1998)
\bibitem{17} Perlmutter, S. et al., ApJ, 517, 565 (1999)
\bibitem{18} Tegmark, M. et al., Phys. Rev. D 69, 103501 (2004)
\bibitem{19} Eisenstein, D.J. et al., ApJ, 633, 560 (2005)
\bibitem{20} Turner, M. \& Huterer, D., astro-ph/07062186
\bibitem{21} Frieman, J.A., Turner, M. and Huterer, D., astro-ph/0803.0982
\bibitem{22} Chongchitnan, S. and Efstathiou, G., Phys. Rev. D 76, 043508 (2007)
\bibitem{23} Weinberg, S. ApJ, 168, 175 (1971)
\end{document}